\documentstyle[spie]{article} 

\newcommand{\bce}{\begin{center}}
\newcommand{\ece}{\end{center}}
\newcommand{\be}{\begin{equation}}
\newcommand{\ee}{\end{equation}}
\newcommand{\bea}{\vspace{0.25cm}\begin{eqnarray}}
\newcommand{\eea}{\end{eqnarray}}

\def\PRL{{Phys. Rev. Lett.} }
\def\PRA{{Phys. Rev.} A }

\title{A new conception experimental test of Bell inequalities 
using non-maximally entangled states} 
\author{M.Genovese\supit{a} , G.Brida\supit{a} , C.Novero\supit{a} and E. Predazzi \supit{b} 
\skiplinehalf 
\supit{a} Istituto Elettrotecnico Nazionale Galileo Ferraris, Str. delle Cacce 91,\\ I-10135 Torino, Italy 
\\
\supit{b}Dip. Fisica Teorica Univ. Torino e INFN, via P. Giuria 1,\\ I-10125 Torino, Italy 
}

\begin{document} 
\maketitle 
\begin{abstract}
We report on a test of Bell inequalities using a non-maximally entangled state, which represents an important step in the direction of eliminating the detection loophole.
The experiment is based on the creation of a polarisation entangled state via the superposition, by use of an appropriate optics, of the spontaneous fluorescence emitted by two non-linear crystals driven by the same pumping laser. The alignment has profitably taken advantage from the use of an optical amplifier scheme, where a solid state laser is injected into the crystals together with the pumping laser.
In principle a very high total quantum efficiency can be reached using this configuration and thus the final version of this experiment can lead to a resolution of the detection loophole, we carefully discuss the conditions which must be satisfied for reaching this result.
\end{abstract}
 
 
\subsection{Introduction} 
In 1935 Einstein-Podolsky-Rosen \cite{EPR}, analysing the measurement on a entangled state, proposed that Quantum Mechanics (QM) could be an incomplete theory, representing a statistical approximation of a complete deterministic theory: this was the born of Local Hidden Variable Theories, where observables values are fixed by some hidden variable and probabilistic predictions become epistemic, being due to our ignorance of hidden variables values. 
A subsequent fundamental progress in discussing Local Hidden Variable (LHV) was Bell's discovery that any theory of this kind must satisfy certain inequalities which can be violated in QM leading in principle to a possible experimental test of the validity of the standard interpretation of QM compared to LHV.
Since then, many interesting experiments have been devoted to a test of Bell inequalities, the most interesting of them using photon pairs \cite{Mandel,asp,franson,type1,type2}, leading to a substantial agreement with quantum mechanics and disfavouring LHV theories. But, up to now, no experiment has yet been able to exclude definitively such theories. In fact, so far, one has always been forced to introduce, at least, a further additional hypothesis \cite{santos}, due to the low total detection efficiency, stating that the observed sample of particle pairs is a faithful subsample of the initial set of pairs. This problem is known as { \it detection or efficiency loophole}. The research for new experimental configurations able to overcome the detection loophole is, of course, of the greatest interest. 
In the 90's a big progress in the direction of eliminating this loophole has been obtained by using parametric down conversion (PDC) processes. 
This technique \cite {Mandel} has been largely employed for generating "entangled" photon pairs, i.e. pairs of photons described by a common wave function which cannot be factorised into the product of two distinct wave functions pertaining to separated photons. 
The generation of entangled states by parametric down conversion (PDC) has replaced other techniques, such as the radiative decay of excited atomic states, as it was in the celebrated experiment of A. Aspect et al. \cite{asp}, for it overcomes some former limitations. In particular, it overcomes the poor angular correlation of atomic cascade photons, that is at the origin of the small total efficiency of this type of experiments in which one is forced to select a small subsample of the produced photons, leading inevitably to the detection loophole, for PDC presents angular correlations better than 1 mrad 
The first experiments using this technique, where performed with type I PDC, which gives phase and momentum entanglement and can be used for a test of Bell inequalities using two spatially separated interferometers \cite{franson}, as realised by Ref.\cite{type1}. The use of beam splitters, however, strongly reduces the total quantum efficiency. 
In alternative, a polarisation entangled state can be generated \cite{ou}. However, in generating this state, in most of the used configurations, half of the initial photon flux is lost and the efficiency loophole cannot be eliminated \cite{santos}. 
Recently, an experiment where a polarisation entangled state is directly generated, has been realised using Type II PDC \cite{type2}. This scheme has permitted, at the price of delicate compensations for having identical arrival time of the ordinary and extraordinary photon, a much higher total efficiency than the previous ones, which is, however, still far from the value of $0.81$ required for eliminating the detection loophole for a maximally entangled state (see later). Also, some recent experiments studying equalities among correlations functions rather than Bell inequalities \cite{dem} are very far from giving a loophole free test of local realism \cite{garuccio}. A large interest remains therefore for new experiments increasing total quantum efficiency in order to reduce and finally overcome the efficiency loophole. 
For this purpose, we have considered \cite{napoli} the possibility of generating a polarisation entangled state via the
superposition of the spontaneous fluorescence emitted by two non-linear crystals (rotated for having orthogonal polarisation) driven
by the same pumping laser \cite{hardy}. The crystals are put in cascade
along the propagation direction of the pumping laser and the superposition
is obtained by using an appropriate optics. If the path between the two
crystal is smaller than the coherence length of the laser, the two photons
path are indistinguishable and a polarisation entangled state is created. In fact, applying the evolution operator given by the PDC Hamiltonian one has, to the first order of the perturbative expansion:
\be
\vert \Psi \rangle =\vert vacuum \rangle + f_1 V_1 \vert H \rangle \vert H \rangle + f_2 V_2 \vert V
\rangle \vert V \rangle 
\label{PsiH}
\ee
where $f_i$ keeps into account the properties of crystal $i$ ($|f_i|^2$ is the fraction of incident light down converted by the non-linear crystal) and $V_i$ the pump intensity at the crystal $i$. 
The possibility of easily obtaining a non maximally entangled state (where $V_1 f_1$
and $V_2 f_2$ are different) is very important, for it has been recognised that for non maximally entangled pairs the lower limit on the total detection efficiency for eliminating the detection loophole is as small as 0.67 \cite{eb} (in the case of no background). This has to be compared with the maximally entangled pairs case, where a total efficiency larger than 0.81 is required. However, it must be noticed that, for non-maximally entangled states, the largest discrepancy between quantum mechanics and local hidden variable theories is reduced: thus a compromise between a lower total efficiency and a still sufficiently large value of this difference will be necessary when realising of an experiment addressed to overcome the detection loophole. 
\subsection{Description of the experiment} 
The general scheme of the experiment is shown in fig. 1: two crystals of $LiIO_3$ (10x10x10 mm) \footnote{Which we have measured to have $d_{31} = 3.5 \pm 0.4$ \cite{BGN}.} are placed along the pump laser propagation, 250 mm apart, a distance smaller than the coherence length of the pumping laser. This guarantees indistinguishability in the creation of a couple of photons in the first or in the second crystal. A couple of planoconvex lenses of 120 mm focal length centred in between focalises the spontaneous emission from the first crystal into the second one maintaining exactly, in principle, the angular spread. A hole of 4 mm diameter is drilled into the centre of the lenses to allow transmission of the pump radiation without absorption and, even more important, without adding stray-light, because of fluorescence and diffusion of the UV radiation. This configuration, which realises a so-called "optical condenser", has been chosen among others, using an optical simulation program, as a compromise between minimisation of aberrations (mainly spherical and chromatic) and losses due to the number of optical components. The pumping beam at the exit of the first crystal is displaced from its input direction by birefringence: the small quartz plate (5 x5 x5 mm) in front of the first lens of the condensers compensates this displacement, so that the input conditions are prepared to be the same for the two crystals, apart from alignment errors. Finally, a half-wavelength plate immediately after the condenser rotates the polarisation of the Argon beam and excites in the second crystal a spontaneous emission cross-polarised with respect to the first one. With a phase matching angle of $51^o$ the spontaneous emissions at 633 and 789 nm (which are the wave lengths used for the test) are emitted at $3.5^o$ and $4^o$ respectively. The dimensions and positions of both plates are carefully chosen in order not to intersect this two stimulated emissions. 
We have used as photo-detectors two avalanche photodiodes with active quenching (EG\&G SPCM-AQ) with a sensitive area of 0.025 $mm^2$ and dark count below 50 counts/s.
PDC signal was coupled to an optical fiber (carrying the light on the detectors) by means of a microscope objective with magnification 20, preceded by a polariser (with extinction ratio $10^{-6}$).
 
The output signals from the detectors are routed to a two channel counter, in order to have the number of events on single channel, and to a 
Time to Amplitude Converter circuit, followed by a single channel analyser, for
selecting and counting coincidence events. 
A very interesting degree of freedom of this configuration is given by the fact that by tuning the pump intensity between the two crystals, one can easily
tune the value of $f=(f_2 V_2)/(f_1 V_1)$, which determines how far from a maximally
entangled state ($f=1$) the produced state is. This is a fundamental property,
which permits to select the most appropriate state for the experiment
The main difficulty of this configuration is in the alignment, which is of fundamental importance for having a high visibility. 
This problem has been solved using a technique, that had been already applied in our laboratory \cite{brida} for metrological studies, namely the use of an optical amplifier
scheme, where a solid state laser is injected into the crystals together with
the pumping laser, an argon laser at 351 nm wavelength (see fig.1). If the angle of injection is selected appropriately, a stimulated emission along the correlated direction appears, permitting to identify quite easily the two correlated directions. Then, stopping the stimulated emission of the first crystal, and rotating the polarisation of the diode laser one obtains the stimulated emission in the second crystal and can check the superposition with the former.
We think that the proposed scheme is well suited for leading to a further step toward a conclusive experimental test of non-locality in quantum mechanics. The
main advantage of the proposed configuration with
respect to most of the previous experimental set-ups is that all the
entangled pairs are selected (and not only $< 50 \%$ as with beams splitters), furthermore it does not require delicate compensations for the optical paths of the ordinary and extraordinary rays after the crystal.
At the moment, the results which we are going to present are still far from
a definitive solution of the detection loophole; nevertheless, being the first test of Bell inequalities using a non-maximally
entangled state, they
represents an important step in this direction. Furthermore, this
configuration permits to use any pair of correlated frequencies and not only
the degenerate ones. We have thus realised this test using for a first time
two different wave lengths (at $633$ and $789$ nm).
It must be acknowledged that a set-up for generating polarisation entangled pairs of photons, which presents analogies with our, has been realised recently in Ref. \cite{Kwiat}
The main difference between the two experiments is that in \cite{Kwiat} the
two crystals are very thin and in contact with orthogonal optical
axes: this permits a "partial" superposition of the two emissions with
opposite polarisation. This overlapping is mainly due to the finite dimension of the pump laser beam, which reflects into a finite width of each wavelength emission.
A much better superposition can be obtained with the present scheme, by fine tuning the crystals' and optics' positions and using the parametric amplifier trick. 
Furthermore, in the experiment of Ref. \cite{Kwiat}
the value of $f$ is in principle tunable by rotating the
polarisation of the pump laser, however this reduces the power of the
pump producing PDC already in the first crystal, while in our case the whole pump
power can always be used in the first crystal, tuning the PDC produced in
the second one. 
More recently \cite{k2}, they have also performed a test about local realism using a particular kind of Hardy equalities \cite{HardyE}. Their result is in agreement with quantum mechanics modulo the detection loophole, no discussion concerning the elimination of loopholes for this equality is presented. 
As a first check of our apparatus, we have measured the interference
fringes, varying the setting of one of the polarisers, leaving the other
fixed. We have found a high visibility, $V=0.973 \pm 0.038$.
Our results are summarised by the value obtained for the Clauser-Horne sum 
\begin{equation}
CH=N(\theta _{1},\theta _{2})-N(\theta _{1},\theta _{2}^{\prime })+N(\theta
_{1}^{\prime },\theta _{2})+N(\theta _{1}^{\prime },\theta _{2}^{\prime
})-N(\theta _{1}^{\prime },\infty )-N(\infty ,\theta _{2}) \label{eq:CH}
\end{equation}
which is strictly negative for local realistic theory. In (\ref{eq:CH}), $N(\theta _{1},\theta _{2})$ is the number of coincidences between
channels 1 and 2 when the two polarisers are rotated to an angle $\theta _{1}$
and $\theta _{2}$ respectively ($\infty $ denotes the absence of selection
of polarisation for that channel).
On the other hand, quantum mechanics predictions for $CH$ can be larger than zero: for a maximally
entangled state the largest value is obtained for $\theta _{1}=67^{o}.5$ , $%
\theta _{2}=45^{o}$, $\theta _{1}^{\prime }=22^{o}.5$ , $\theta _{2}^{\prime
}=0^{o}$ and corresponds to a ratio 
\begin{equation}
R=[N(\theta _{1},\theta _{2})-N(\theta _{1},\theta _{2}^{\prime })+N(\theta
_{1}^{\prime },\theta _{2})+N(\theta _{1}^{\prime },\theta _{2}^{\prime
})]/[N(\theta _{1}^{\prime },\infty )+N(\infty ,\theta _{2})] \label{eq:R}
\end{equation}
equal to 1.207.
For non-maximally entangled states the angles for which CH is maximal are
somehow different and the maximum is reduced to a smaller value. The angles corresponding to
the maximum can be evaluated maximising Eq. \ref{eq:CH} with 
\bea
\left. \begin{array}{l}
N[\theta _{1},\theta _{2}] = [ \epsilon _1^{||} \epsilon _2^{||} (Sin[\theta _{1}]^{2}\cdot Sin[\theta_{2}]^{2}) + \\
\epsilon _1^{\perp} \epsilon _2^{\perp} 
(Cos[\theta _{1}]^{2} \cdot Cos[\theta _{2}]^{2} )\\
(\epsilon _1^{\perp} \epsilon _2^{||} Sin[\theta _{1}]^2\cdot Cos[\theta _{2}]^2 + \epsilon _1^{||} \epsilon _2^{\perp} 
Cos[\theta _{1}]^2 \cdot Sin [\theta _{2}]^2 ) \\
+ |f|^{2}\ast (\epsilon _1^{\perp} \epsilon _2^{\perp} (Sin[\theta _{1}]^{2}\cdot Sin[\theta_{2}]^{2}) + \epsilon _1^{||} \epsilon _2^{||}
(Cos[\theta _{1}]^{2} Cos[\theta _{2}]^{2} ) +\\
(\epsilon _1^{||} \epsilon _2^{\perp} Sin[\theta _{1}]^2\cdot Cos[\theta _{2}]^{2} +\\
\epsilon _1^{\perp} \epsilon _2^{||} 
Cos[\theta _{1}]^2 \cdot Sin [\theta _{2}]^2 ) \\
+ (f+f^{\ast }) ((\epsilon _1^{||} \epsilon _2^{||} + \epsilon _1^{\perp} \epsilon _2^{\perp} - \epsilon _1^{||} \epsilon _2^{\perp} - 
\epsilon _1^{\perp} \epsilon _2^{||})
\cdot (Sin[\theta _{1}]\cdot Sin[\theta _{2}]\cdot
Cos[\theta _{1}]\cdot Cos[\theta _{2}]) ] /(1+|f|^{2}) \,
\end{array}\right. \, . 
\label{cc}
\eea
where (for the case of non-ideal polariser) $\epsilon _i^{||}$ and $\epsilon _i^{\perp}$ 
correspond to the transmission when the polariser (on the branch $i$) axis is aligned or normal to the polarisation axis respectively.
The limits for obtaining a detection loophole free experiment using non-maximally entangled states with non-ideal polarisers are summarised in fig.2, where one has maximised the Clauser-Horne inequality violation using the correlation function (\ref{cc}) for the non-maximally entangled case. From the figure one derives that for the ideal polariser case one can lower the limit on the total efficiency up to 0.67 for a opportune non-maximally entangled state. The use of non ideal polarisers slightly enhances this limit, however even for a extinction ratio ($
\epsilon ^{||} \cdot \epsilon ^{\perp} $) 
of $10^{-4}$ the effect is small (see fig. 2). Considering the extinction ratios of commercial polarisers, this simulation shows that this does not represent a significant problem. Anyway, having $\epsilon ^{||} < 1$ contributes to lower the detection efficiency. An important factor is the background on the single channel detection: its effect is exactly the same of a lower quantum efficiency of photodetection, for it increases the denominator in Eq. \ref{eq:R}.
The phase of $f$ must be kept next to zero. Any relative phase between the two components of the entangled state reflects into a reduction of Clauser-Horne inequality violation, up to reaching no violation at all for a phase difference of $\pi /2$. 
In our case we have generated a state with $f \simeq 0.4$: in this case the largest violation of the inequality is reached for $\theta_1 =72^o.24$ , $\theta_2=45^o$, $\theta_1 ^{\prime}= 17^o.76$ and 
$\theta_2 ^{\prime}= 0^o$, to $R=1.16$. Our experimental result is $CH = 512 \pm 135$ coincidences per second, 
which is almost four standard deviations different from zero and compatible with the theoretical value predicted by quantum mechanics.
In terms of the ratio (\ref{eq:R}), our result is $1.082 \pm 0.031$.
For the sake of comparison, one can consider the value obtained with the angles which optimise Bell inequalities violation for a maximally entangled state.
The result is $CH = 92 \pm 89$, which, as expected, shows a smaller violation than the value obtained with the correct angles setting.
\subsection{Conclusions} 
We have presented the first measurement of the violation of Clauser-Horne inequality (or of other Bell inequalities) using a non-maximally entangled state. This represents a relevant step in the direction of eliminating the detection loophole. Further developments in this sense are the purpose of this collaboration.
Besides, this scheme gives a beautiful example of how entanglement and quantum interference derive from absence of welcher weg (which path) information \cite{eraser}, namely from the impossibility of distinguishing in which crystal the pair is produced. In fact, this scheme works substantially as a quantum eraser: when no polariser is inserted after the
crystals, interference is cancelled because of different polarisations produced in the two crystals, when a polariser, rotated respect to vertical (horizontal) axis,
is inserted, due to (at least partial) cancellation of the information about where PDC has happened interference is restored (and could be modulated changing the phase between the pair produced in the first and second crystal).

\subsection{Acknowledgments} 
We would like to
acknowledge the support of the Italian Space Agency under contract LONO 500172 and
of MURST via special programs "giovani ricercatori", Dip. Fisica Teorica Univ. Torino. 
\newpage
{\bf Figures captions}
- Fig. 1 Sketch of the source of polarisation entangled photons. CR1 and CR2 are two $LiIO_3$ crystals cut at the phase-matching angle of $51^o$. L1 and L2 are two identical piano-convex lenses with a hole of 4 mm in the centre. P is a 5 x 5 x 5 mm quartz plate for birefringence compensation and $\lambda / 2$ is a first order half wave-length plate at 351 nm. 
U.V. identifies the pumping radiation at 351 nm. The infrared beam (I.R.) at 789 nm is generated by a diode laser and is used for system alignment only. The parametric amplifier scheme, described in the text, is shown as well.
The dashed line identifies the idler radiation at 633 nm. A second half-wave plate on the I.R. beam (not shown in the figure) allows amplified idler emission from the second crystals too. The figure is not in scale.
- Fig. 2 a) Contour plot of the quantity $CH/N$ (see Eq. \ref{eq:CH}. N is the total number of detections) in the plane with $f$ (non maximally entanglement parameter, see the text for the definition) as y-axis and $\eta$ (total detection efficiency) as x-axis. The polarisers are supposed to have $\epsilon _i^{||}=1$. The leftmost region corresponds to the region where no detection loophole free test of Bell inequalities can be performed. The contour lines are at 0, 0.01, 0.1, 0.15, 0.2.
b) the same as (a), but with an extinction ratio $10^{-4}$.

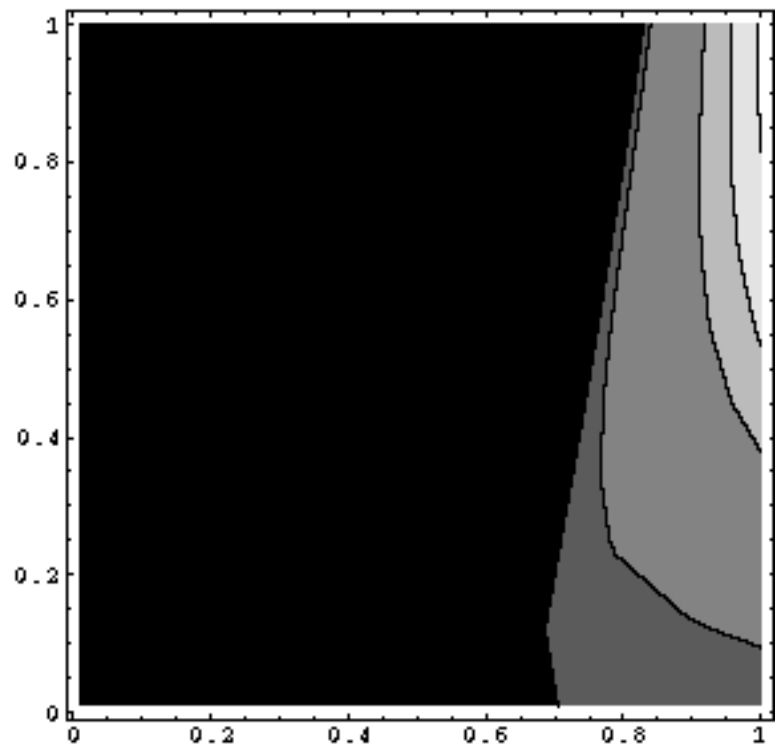
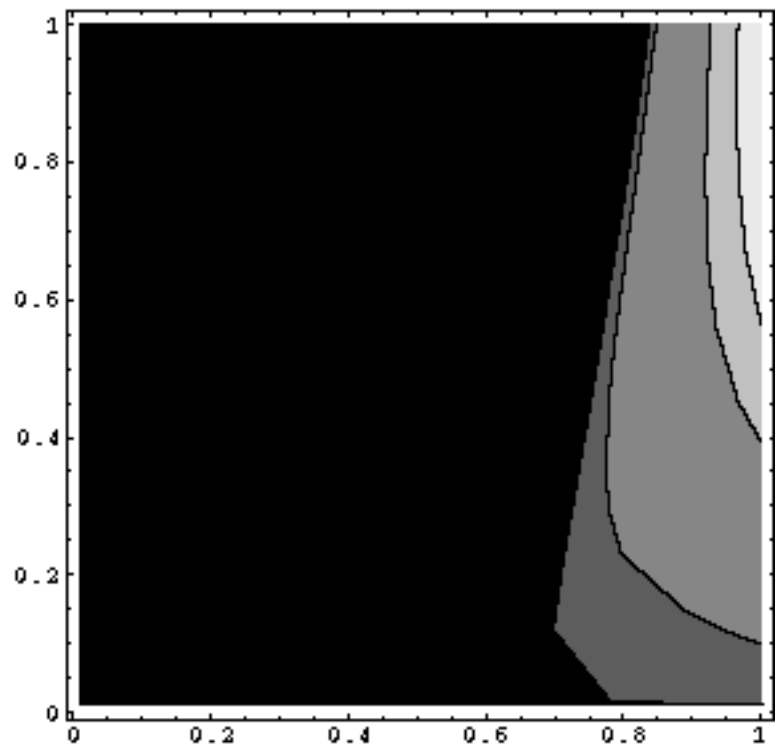
\end{document}